\documentstyle[preprint,aps,epsf,tighten]{revtex}       % PREPRINT STYLE

\begin{document}
%\draft
%
%%%%%%%%%%%%%%%%%%%%% TITLE PAGE %%%%%%%%%%%%%%%%%%%%%%%%%%%%%%%%%%%%%%%%%
%
\preprint{$
\begin{array}{l}
\mbox{BA-00-30}\\
\mbox{FERMILAB-Conf-00/130-T}\\
%\mbox{hep--ph/000}\\
\mbox{June 2000}\\[0.5in]
\end{array}
$}
\title{EXPLICIT SO(10) SUPERSYMMETRIC\\
	GRAND UNIFIED MODEL\footnote{Paper contributed to the NEUTRINO 2000
	Conference in Sudbury, Canada, 16-21 June 2000\\
	\hspace*{0.15in}and presented at the SUSY2K Conference at CERN, 1 July 
	2000}\\[0.5in]}
\author{Carl H. Albright$^1$ and S.M. Barr$^2$}
\address{$^1$Department of Physics, Northern Illinois University, DeKalb, 
        IL 60115\\
        and \\
        Fermi National Accelerator Laboratory, P.O. Box 500, Batavia, IL 
        60510\\
        $^2$Bartol Research Institute, University of Delaware,
        Newark, DE 19716}
\maketitle
\begin{abstract}
A complete set of Higgs and matter superfields is introduced with well-defined
$SO(10)$ properties and $U(1) \times Z_2 \times Z_2$ family charges from 
which the Higgs and Yukawa superpotentials are constructed.  The Higgs 
fields solve the doublet-triplet splitting problem, while the structures
of the four Dirac fermion mass matrices obtained involve just six 
effective Yukawa operators.  The right-handed Majorana matrix, $M_R$, arises 
from one Higgs field coupling to several pairs of superheavy conjugate 
neutrino singlets.  In terms of 10 input parameters to the mass matrices,
the model accurately yields the 20 masses and mixings of the lightest
quarks and leptons, as well as the masses of the 3 heavy right-handed
neutrinos.  The bimaximal atmospheric and solar neutrino vacuum solutions
are favored in this simplest version with a moderate hierarchy in $M_R$.
The large mixing angle MSW solution is obtainable, on the other hand, with a 
considerably larger hierarchy in $M_R$ which is also necessary to obtain 
baryogenesis through the leptogenesis mechanism.\\[0.5in]
\thispagestyle{empty}
\end{abstract}
%
%\vskip .2in
%
\newpage
This conference paper presents a brief summary of recent work \cite{ab6},
\cite{ab7} carried out by the authors, where it is shown how one can start
from a set of Higgs and matter superfields specified by their $SO(10)$
properties and $U(1) \times Z_2 \times Z_2$ family charges, construct the 
Higgs and Yukawa superpotentials, and derive fermion mass and mixing matrices
for both quarks and leptons.  Previously in several publications \cite{ab1}
- \cite{ab5}, an effective 
operator approach involving the same symmetries was employed to postulate 
Dirac mass matrices with similar textures which lead to accurate predictions
for the quark and charged lepton masses and CKM mixing matrix.  But the 
right-handed neutrino Majorana matrix remained undetermined.  With our more
fundamental approach discussed here, the simplest input version favors 
the bimaximal atmospheric and solar neutrino vacuum solutions.

In the minimal $SO(10)$ grand unified theory, the two Higgs doublets which
break the electroweak symmetry lie completely in one vector representation,
${\bf 10}_H$, which contains a $5$ and $\bar{5}$ of $SU(5)$: 
$H_U \subset 5({\bf 10}_H),\quad H_D \subset \bar{5}({\bf 10}_H)$.  The 
colored Higgs triplets also present in the ${\bf 10}_H$ can be made 
superheavy at the GUT scale, $\Lambda_G$, by means of the Dimopoulos-Wilczek 
mechanism \cite{dim-wil}, in the presence of a Higgs adjoint ${\bf 45}_H$ 
whose VEV points in the $B - L$ direction and a second ${\bf 10'}_H$ which gets 
massive.  Barr and Raby \cite{b-r} showed that this solution of the 
doublet-triplet splitting problem can be stabilized with the introduction 
of two pairs of spinor $({\bf 16}_H,\ {\bf \overline{16}}_H)$'s plus 
several Higgs singlets.  One pair gets $SU(5)$-singlet VEV's at $\Lambda_G$ 
and together with the ${\bf 45}_H$ breaks $SO(10)$ down to the standard
model.  One can then arrange the surviving Higgs doublets to be 
\begin{equation}
	H_U \subset 5({\bf 10}_H),\quad 
  	H_D \subset \bar{5}({\bf 10}_H)\cos \gamma + \bar{5}({\bf 16'}_H)
	\sin \gamma
\end{equation}
while the combination orthogonal to $H_D$ gets massive at the GUT scale.
With both the $\bar{5}({\bf 10}_H)$ and the $\bar{5}({\bf 16'}_H)$ getting
a VEV at the EW scale, complete $t - b - \tau$ Yukawa coupling unification
is possible with $\tan \beta \equiv v_u/v_d \ll 55$.  This minimal $SO(10)$
Higgs structure can naturally be achieved with a global $U(1) \times 
Z_2 \times Z_2$ symmetry.  

With the above construction as a natural starting point for our model,
we add several more Higgs ${\bf 10}$'s and singlets in order to construct
the full Higgs superpotential necessary for the generation of the desired
fermion mass matrices.  A complete listing of the Higgs superfields is 
given in Table I.  
$$\begin{tabular}{ll}
\multicolumn{2}{l}{\bf Higgs\ Fields\ Needed\ to\ Solve\ 
        the\ 2-3\ Problem:}\\[0.1in]
        ${\bf 45}_{B-L}$: & $A(0)^{+-}$ \\
        ${\bf 16}$: & $C(\frac{3}{2})^{-+},\ 
                C'(\frac{3}{2}-p)^{++}$\\
        $\overline{\bf 16}$: & $\bar{C}(-\frac{3}{2})^{++},
                \ \bar{C}'(-\frac{3}{2}-p)^{-+}$\\
        ${\bf 10}$: & $T_1(1)^{++},\ T_2(-1)^{+-}$\\
        ${\bf 1}$: & $X(0)^{++},\ P(p)^{+-},\ Z_1(p)^{++},
                \ Z_2(p)^{++}$\\[0.1in]
\multicolumn{2}{l}{\bf Additional Higgs\ Fields\ for\ the\ Mass\ Matrices:}
        \\[0.1in]
        ${\bf 10}$: & $T_0(1+p)^{+-},\ T'_o(1+2p)^{+-}$,\\
                & \ $\bar{T}_o(-3+p)^{-+},\ \bar{T}'_o(-1-3p)^{-+}$\\
        ${\bf 1}$: & $Y(2)^{-+},\ Y'(2)^{++},\ 
                S(2-2p)^{--},\ S'(2-3p)^{--}$,\\
                & $V_M(4+2p)^{++}$\\[0.1in]
\multicolumn{2}{c}{\bf Table I.  Higgs superfields in the proposed model.}
\end{tabular}$$
The full Higgs superpotential follows from the $SO(10)$ and $U(1) \times
Z_2 \times Z_2$ assignments:
\begin{equation}
\begin{array}{rcl} 
        W_{\rm Higgs} &=& W_A + W_{CA} + W_{2/3} + W_{H_D} + W_R\\[0.1in]
        W_A &=& tr A^4/M + M_A tr A^2\\[0.1in]
        W_{CA} &=& X(\overline{C}C)^2/M^2_C + F(X) \\
         & & + \overline{C}'(PA/M_1 + Z_1)C + \overline{C}(PA/M_2 + 
          Z_2)C'\\[0.1in]
        W_{2/3} &=& T_1 A T_2 + Y' T^2_2\\[0.1in]
        W_{H_D} &=& T_1 \overline{C}\overline{C} Y'/M + 
          \overline{T}_0 C C' + \overline{T}_0(T_0 S + T'_0 S')\\[0.1in]
        W_R &=& \overline{T}_0 \overline{T}'_0 V_M\\
\end{array}
\end{equation}

\noindent  The properties described above are obtained from a study of the 
F-flat constraint conditions, while the role of the extra Higgs fields 
will become apparent in Eq. (3).

The matter superfields consist of three chiral families together with 
two vector-like pairs of spinor fields, one pair of vector fields and 
three pairs of singlets.  All but the chiral fields are integrated out
to obtain the GUT scale structure for the fermion mass matrices.  A 
complete listing is given in Table II.\\[0.2in]
$$\begin{tabular}{lll}
        ${\bf 16}_1(-\frac{1}{2}-2p)^{+-}$ \  & 
                ${\bf 16}_2(-\frac{1}{2}+p)^{++}$ \ & 
                ${\bf 16}_3(-\frac{1}{2})^{++}$ \\
        ${\bf 16}(-\frac{1}{2}-p)^{-+}$ \ & 
                ${\bf 16}'(-\frac{1}{2})^{-+}$ \\
        $\overline{\bf 16}(\frac{1}{2})^{+-}$ \ &
        $\overline{\bf 16}'(-\frac{3}{2}+2p)^{+-}$\\[0.1in]
        ${\bf 10}_1(-1-p)^{-+}$ \ & ${\bf 10}_2(-1+p)^{++}$
          \\[0.1in]
        ${\bf 1}_1(2+2p)^{+-}$ \ & 
                ${\bf 1}_2(2-p)^{++}$ \  
                & ${\bf 1}_3(2)^{++}$\\[0.1in]
        ${\bf 1}^c_1(-2-2p)^{+-}$ \ & ${\bf 1}^c_2(-2)^{+-}$
                \ & ${\bf 1}^c_3(-2-p)^{++}$ \\[0.1in]
\multicolumn{3}{c}{\bf Table II.  Matter superfields in the proposed
model.}\\[0.2in]
  \end{tabular}$$

The Yukawa superpotential can then be uniquely constructed from the Higgs 
and matter superfields by taking into account their $SO(10)$ properties 
and $U(1) \times Z_2 \times Z_2$ family charges.  One finds
\begin{equation}
\begin{array}{rl}
        W_{Yukawa} &= {\bf 16}_3 \cdot {\bf 16}_3 \cdot T_1 + {\bf 16}_2
                \cdot {\bf 16} \cdot T_1
                + {\bf 16}' \cdot {\bf 16}' \cdot T_1\\
                &+ {\bf 16}_3 \cdot {\bf 16}_1 \cdot T'_0
                + {\bf 16}_2 \cdot {\bf 16}_1 \cdot T_0
                + {\bf 16}_3 \cdot {\overline{\bf 16}} \cdot A\\
                &+ {\bf 16}_1 \cdot \overline{\bf 16}' \cdot Y'
                + {\bf 16} \cdot {\overline{\bf 16}} \cdot P
                + {\bf 16}' \cdot {\overline{\bf 16}}' \cdot S\\
                &+ {\bf 16}_3 \cdot {\bf 10}_2 \cdot C'
                + {\bf 16}_2 \cdot {\bf 10}_1 \cdot C
                + {\bf 10}_1 \cdot {\bf 10}_2 \cdot Y\\
                &+ {\bf 16}_3 \cdot {\bf 1}_3 \cdot \overline{C}
                + {\bf 16}_2 \cdot {\bf 1}_2 \cdot \overline{C}
                + {\bf 16}_1 \cdot {\bf 1}_1 \cdot \overline{C}\\
                &+ {\bf 1}_3 \cdot {\bf 1}^c_3 \cdot Z
                + {\bf 1}_2 \cdot {\bf 1}^c_2 \cdot P
                + {\bf 1}_1 \cdot {\bf 1}^c_1 \cdot X\\
                &+ {\bf 1}^c_3 \cdot {\bf 1}^c_3 \cdot V_M
                + {\bf 1}^c_1 \cdot {\bf 1}^c_2 \cdot V_M\\
  \end{array}
\end{equation}

\noindent  Note that the right-handed Majorana matrix elements are all
generated through the Majorana couplings of the $V_M$ Higgs field with 
the conjugate fermions listed in Table II.

In order to derive the four Dirac mass matrices $U,\ D,\ N,\ L$, from Eq. (3)
one first finds the three zero mass eigenstates for each type of fermion 
$f = u_L,\ u^c_L,\ d_L,\ d^c_L,\ \ell^-_L,\ \ell^+_L,\ \nu_L$ and $\nu^c_L$
for the superheavy mass matrix connecting $f$ to $\bar{f}$ at $\Lambda_G$
with the electroweak VEV's set equal to zero.  The terms in Eq. (3) 
involving $\langle T_1 \rangle,\ \langle C' \rangle,\ \langle T_0 \rangle$, 
and $\langle T'_0 \rangle$ then give rise to the $3 \times 3$ Dirac 
mass matrices coupling $u_L$ to $u^c_L$, etc.  See \cite{ab6} for details.
Just 6 effective Yukawa operators are obtained of the following types 
connecting the two families indicated:
\begin{equation}
\begin{array}{rl}
   33:& {\bf 16}_3 \cdot {\bf 10}_H \cdot {\bf 16_3}\\[0.05in]
   23:& \left[{\bf 16}_2 \cdot {\bf 10}_H\right]_{16} \cdot
        _{\overline{16}}\left[{\bf 45}_H \cdot {\bf 16}_3\right]
        /M_G\\[0.05in]
   23:& \left[{\bf 16}_2 \cdot {\bf 16}_H\right]_{10_1} \cdot
        _{10_2}\left[{\bf 16'}_H \cdot {\bf 16}_3\right]/M_G\\[0.05in]
   13:& \left[{\bf 16}_1 \cdot {\bf 16}_3\right]_{10''_H} \cdot
        _{\overline{10}'_H}\left[{\bf 16}_H \cdot {\bf 16'}_H\right]
        /M_G\\[0.05in]
   12:& \left[{\bf 16}_1 \cdot {\bf 16}_2\right]_{10'_H} \cdot
        _{\overline{10}'_H}\left[{\bf 16}_H \cdot {\bf 16'}_H\right]
        /M_G\\[0.05in]
   11:& \left[{\bf 16}_1 \cdot {\bf 1}_H\right]_{\overline{16}'}
        \cdot\ _{16'}\left[{\bf 10}_H\right]_{16'} \cdot
        _{\overline{16}'}\left[{\bf 1}_H \cdot {\bf 16}_1\right]
        /M^2_G\\
\end{array}
\end{equation}

\noindent The subscripts on the bracketed terms indicate the fermion
or Higgs fields which are contracted with a Higgs scalar and integrated
out.  The $SU(5)$ structure of the above operators will determined
which of the matrix elements actually appear in $U,\ D,\ N$ and $L$.
Note that only the 33 operator is renormalizable, while the 23, 13 and 12 
operators are dimension-5 and the 11 operator is dimension-6.  

The Dirac matrix elements determined by these effective operators are 
conveniently 
pictured in terms of Froggatt-Nielsen diagrams \cite{f-n} in Figs. 1 and 2, 
where the fields integrated out are explicitly given.  The convention
is that the $f^c_{iL}$ lines for the light fermions appear on the left,
while the $f_{iL}$ lines appear on the right.  The first 23 operator involving
${\bf 45}_H$ contributes in an antisymmetric fashion to the 23 and 32 
matrix elements and is proportional to the $B - L$ quantum number of the 
fermion line involved.  The $SU(5)$ structure of the second 23 operator 
reveals a 23 element contribution only to the $D$ matrix and a 32 element
contribution only to the $L$ matrix.

The textures for the four Dirac mass matrices arise from the above effective
operators and diagrammatic structures and can be parametrized as follows:

\begin{equation}
\begin{array}{ll}
U = \left( \begin{array}{ccc} \eta & 0 & 0 \\
  0 & 0 & \epsilon/3 \\ 0 & - \epsilon/3 & 1 \end{array} \right), 
  & D = \left( \begin{array}{ccc} 0 & \delta & \delta' e^{i\phi}
  \\
  \delta & 0 & \sigma + \epsilon/3  \\
  \delta' e^{i \phi} & - \epsilon/3 & 1 \end{array} \right), \\ & \\
N = \left( \begin{array}{ccc} \eta & 0 & 0 \\ 0 & 0 & - \epsilon \\
  0 & \epsilon & 1 \end{array} \right),
  & L = \left( \begin{array}{ccc} 0 & \delta & \delta' e^{i \phi} \\
  \delta & 0 & -\epsilon \\ \delta' e^{i\phi} & 
  \sigma + \epsilon & 1 \end{array} \right),
\end{array}
\end{equation}

\noindent With $\sigma \gg \epsilon$, the Georgi-Jarlskog relations \cite{g-j}
are recovered:

\begin{equation}
	m_s/m_b = m_\mu/3m_\tau,\qquad m_d/m_s = 3 m_e/m_\mu
\end{equation}

\noindent The asymmetrical ``lopsided'' $\sigma$ entries appearing in $D$ 
and $L$ arise from the $SU(5)$ structure of the $\langle \bar{5}({\bf 16}_H) 
\rangle$ VEV noted above.  They can account for the small $V_{cb}$ quark 
mixing and the large atmospheric $\nu_\mu - \nu_\tau$ mixing \cite{atm}, 
since a small left-handed 
rotation in the 23 sector is required to diagonalize $D$ while a large 
left-handed rotation is required to diagonalize $L$ as demonstrated later.
A more detailed study given in \cite{ab7} reveals that the above simple 
textures are obtained provided $\tan \beta$ is not too large: $\tan \beta 
\le 5 - 10$.

The right-handed Majorana mass matrix $M_R$ can be constructed in a similar
fashion and involves the Majorana Higgs $V_M$ coupling massive singlet 
conjugate neutrinos.  The effective dimension-6 operators involved are 

\begin{equation}
  (M_R)_{ij}:\ \left[{\bf 16}_i \cdot {\bf \overline{16}}_H\right]
             _{1_i} \cdot _{1^c_i}\left[{\bf V_M}\right]_{1^c_j} \cdot
             _{1_j}\left[{\bf {\overline{16}}_H \cdot {\bf 16}_j}\right]
             /M^2_G,\quad ij = 33, 12
\end{equation}

\noindent  The corresponding Froggatt-Nielsen diagrams for the 33, 12 and 
21 elements of $M_R$ are shown in Fig. 3.  The matrix can be parameterized
by

\begin{equation}
	M_R = \left(\matrix{ 0 & A\epsilon^3 & 0 \cr A\epsilon^3 & 0 & 0 \cr 
        0 & 0 & 1 \cr}\right)\Lambda_R
\end{equation}

\noindent The effective light neutrino mass matrix follows according to 

\begin{equation}
	M_\nu = N^T M_R^{-1} N = \left(\matrix{ 0 & 0 & 
        	-\eta/(A\epsilon^2)\cr
        	0 & \epsilon^2 & \epsilon\cr 
        	-\eta/(A\epsilon^2) & \epsilon & 1\cr}\right)
        	M^2_U/\Lambda_R
\end{equation}

\noindent  A rotation in the 2-3 plane by an angle $\epsilon$ followed by
a rotation in the 1-3 plane by an angle $(\eta/A)\epsilon$ brings $M_\nu$
into an apparent pseudo-Dirac form whereby two of the neutrinos are 
nearly degenerate, provided $|\eta/A| \ll 1$.  This result corresponds 
most readily to the vacuum ``just-so'' solar neutrino solution \cite{vac}. 

The masses and mixings of the quarks and leptons follow by diagonalization
of the matrices in Eqs. (5) and (9), with the mixing matrices at the GUT
scale given by 

\begin{equation}
	V_{CKM} = U^\dagger_U U_D,\qquad U_{MNS} = U^\dagger_\nu U_L
\end{equation}

\noindent In order to obtain numerical results for the predictions at 
the low scales, the masses and mixings at $\Lambda_G$ were evolved from
the GUT scale to the SUSY scale by use of the 2-loop MSSM beta functions
and then to the running mass or 1 GeV scale with the 3-loop QCD and 1-loop
QED renormalization group equations.  We have used

\begin{equation}
  \begin{array}{rlrlrl}
        \tan \beta =&5, & \Lambda_G =&2 \times 10^{16}
        \ {\rm GeV},\quad & \Lambda_{SUSY} =&m_t(m_t)\\[0.1in]
        \alpha_s(M_Z) =&0.118,\quad & \alpha(M_Z) =&1/127.9, & 
	\sin^2 \theta_W =&0.2315\\
  \end{array}
\end{equation}

The 10 model mass matrix parameters are chosen to be 

\begin{equation}
  \begin{array}{rlrlrlrl}
        \sigma =&1.78, & \epsilon =&0.145, & \eta =&\multicolumn{2}{l}{8 
        \times 10^{-6}}\\[0.05in]
        \delta =&0.0086, & \delta' =&0.0079, & \phi =&54^o, & A =&0.05\\[0.05in]
        M_U =&113\ {\rm GeV}, & M_D =&1\ {\rm GeV}, & \Lambda 
        =&\multicolumn{2}{l}{2.4 \times 10^{14}\ {\rm GeV}}\\
  \end{array}
\end{equation}

\noindent which lead to the following 10 input observables \cite{data}:

\begin{equation}
  \begin{array}{rlrl}
        m_t(m_t) =&165\ {\rm GeV}, & m_u(1\ {\rm GeV}) =&4.5\ 
                {\rm MeV}\\[0.05in]
        m_\tau =&1.777\ {\rm GeV}, & m_\mu =&105.7\ {\rm MeV}\\[0.05in] 
        m_e =&0.511\ {\rm MeV}, & V_{us} =&0.220\\[0.05in]
        V_{cb} =&0.0395, & \delta_{CP} =&64^o\\[0.05in]
        m_3 =&54.3\ {\rm meV}, & \ m_2 =&59.6\ {\rm \mu eV}\\[0.1in]
  \end{array}
\end{equation}

\noindent  The following 5 predictions are then obtained for the 
remaining quark observables:

\begin{equation}
  \begin{array}{rlrl}
	m_c(m_c) =&1.23\ {\rm GeV}, &
                m_b(m_b) =&4.25\ {\rm GeV}\\[0.05in]
        m_s({\rm 1\ GeV}) =&148\ {\rm MeV}, &
                m_d({\rm 1\ GeV}) =&7.9\ {\rm MeV}\\[0.05in]
        |V_{ub}/V_{cb}| =&0.080\\
  \end{array}
\end{equation}

\noindent The good agreement with the experimental quark data is shown
in Fig. 4 for the CKM unitarity triangle in effectively the $\tilde{\rho}
- \tilde{\eta}$ plane.  

The 5 additional predictions for the lepton observables are

\begin{equation}
  \begin{array}{rlrl}
 	m_1 =&56.5\ {\rm \mu eV}\\[0.05in] 
        U_{e2} =&0.733, & U_{\mu 3} =&-0.818\\[0.05in]
        U_{e3} =&0.047, & \delta'_{CP} =& -0.2^o\\
  \end{array}
\end{equation}

\noindent from which one finds the neutrino oscillation solutions

\begin{equation}
  \begin{array}{ll}
        \Delta m^2_{23} = 3.0 \times 10^{-3}\ {\rm eV^2},\quad &        
                \sin^2 2\theta_{atm} = 0.89\\[0.05in]
        \Delta m^2_{12} = 3.6 \times 10^{-10}\ {\rm eV^2},\quad &
                \sin^2 2\theta_{solar} = 0.99\\[0.05in]
        \Delta m^2_{13} = 3.0 \times 10^{-3}\ {\rm eV^2},\quad &
                \sin^2 2\theta_{reac} = 0.009\\
  \end{array}
\end{equation}

\noindent  Finally the 3 heavy right-handed Majorana masses are 
determined to be 

\begin{equation}
	M_3 = 2.4 \times 10^{14}\ {\rm GeV},\quad M_2 = M_1 = 3.66 \times
		10^{10}\ {\rm GeV}
\end{equation}

It is clear that maximal atmospheric neutrino mixing arises from
the structure of charged lepton mass matrix $L$, while the maximal 
mixing vacuum ``just-so'' solar neutrino solution emerges from the very
simple form assumed for the coupling of the $V_M$ Higgs singlet with 
the conjugate neutrino states in the superpotential.  
It also depends critically on the appearance of the small parameter 
$\eta$ appearing in the Dirac matrix $N$, corresponding to the non-zero
$\eta$ entry in $U$ resulting in a non-zero up quark mass at the GUT scale.
The bimaximal large mixing angle MSW solution \cite{MSW} requires a 
considerably larger hierarchy in $M_R$.  But such a large hierarchy may be
desirable in order to obtain baryogensis through the leptogenesis 
mechanism \cite{rm}.  Other more complex textures resulting from a larger 
set of matter superfields and a more complex Yukawa superpotential are 
possible but require one or more additional model parameters and are hence 
less predictive.

The result for $\delta'_{CP}$ obtained above indicates that CP violation 
in the leptonic sector is predicted to be negligible.  On the other hand,
while $\sin^2 2\theta_{reac} = 0.009$ is considerably below the present 
CHOOZ reactor bound of 0.1 \cite{CHOOZ}, it should be measurable at a 
neutrino factory.

Many additional details of the model described are given in references
\cite{ab6} and \cite{ab7}.\\[0.3in]

The research of SMB was supported in part by the Department of Energy under 
contract No. DE-FG02-91ER-40626.  Fermilab is operated by Universities 
Research Association Inc. under contract No. DE-AC02-76CH03000 with the 
Department of Energy.\\[-0.4in]

%
%
%%%%%%%%%%%%%%%%%%%%% REFERENCES %%%%%%%%%%%%%%%%%%%%%%%%%%%%%%%%%%%%%%%%%
%

%
%%%%%%%%%%%%%%%%%%%%%% FIGURE CAPTIONS %%%%%%%%%%%%%%%%%%%%%%%%%%%%%%%%%%
%
% FIG. 1
\begin{figure}
\vspace*{-0.7in}
\centerline{
\epsfxsize=1\hsize
\epsfbox{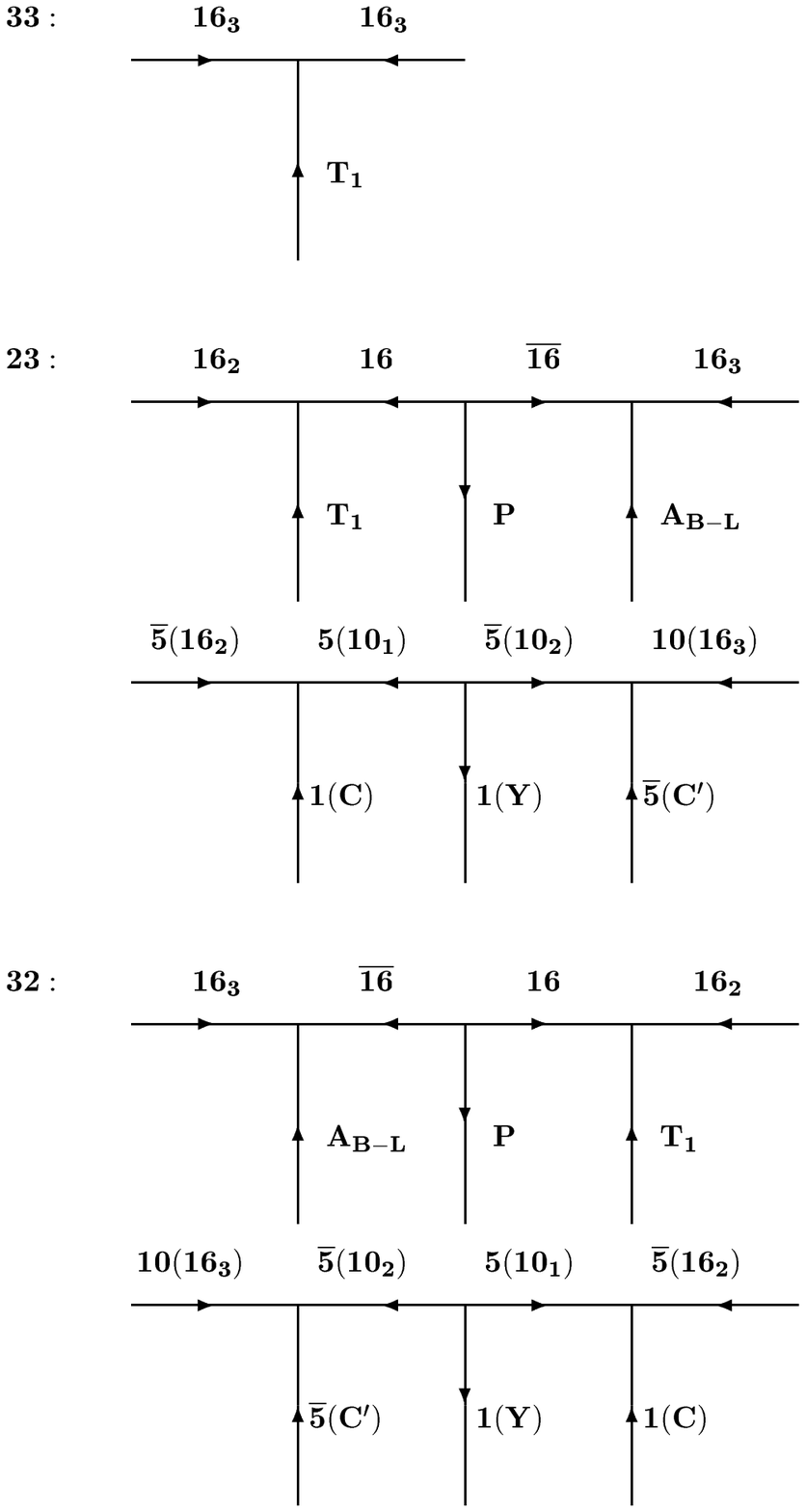}
}
\vspace {0.75in}
\caption{Froggatt-Nielsen diagrams that generate the 33, 23 and 32 
	elements of the quark and 
	\hspace*{0.80in}lepton Dirac mass matrices.  The second 23
	diagram contributes only to $D_{23}$, while 
	\hspace*{0.80in}the second 32
	diagram contributes only to $L_{32}$.}
\end{figure}
\newpage
% FIG. 2
\begin{figure}
\phantom{x}
\centerline{
\epsfxsize=1\hsize
\epsfbox{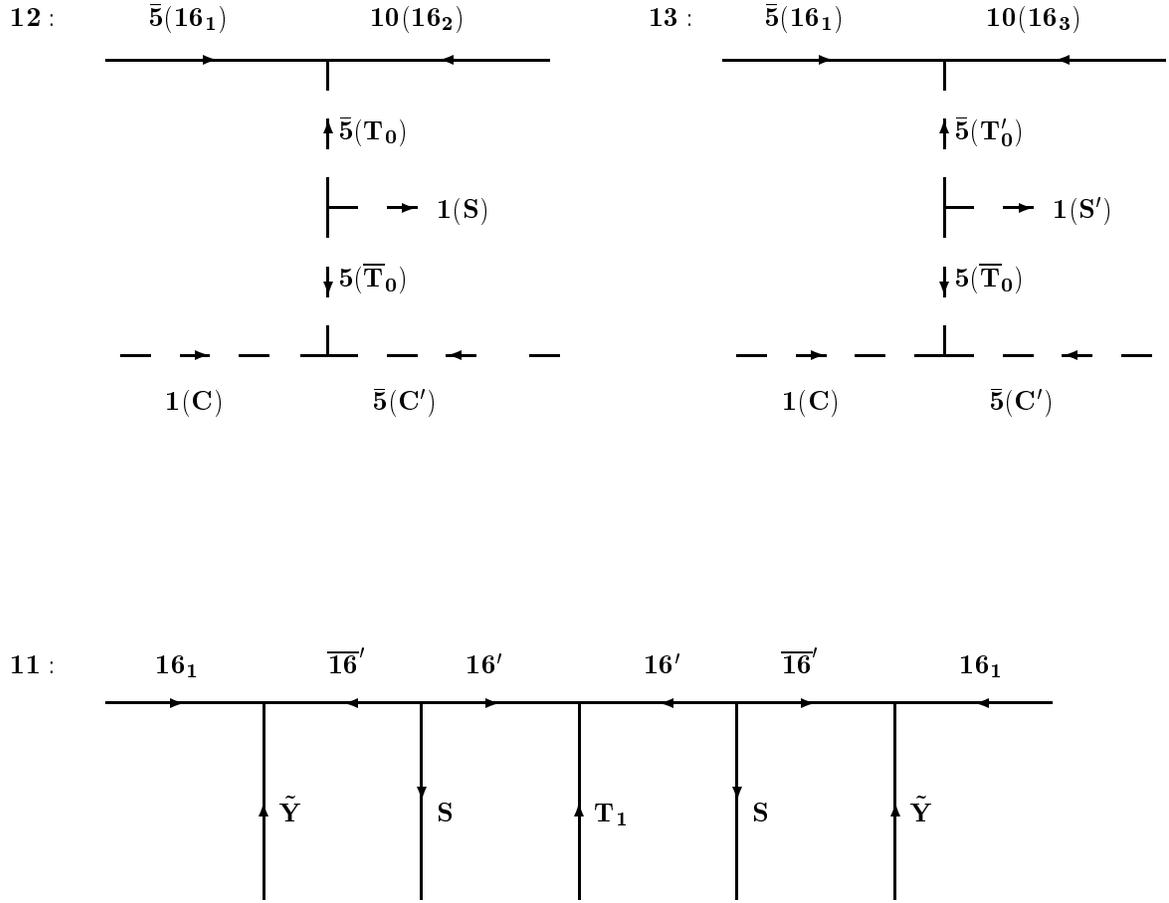}
}
\vspace{0.7in}
\caption{Additional diagrams for the symmetrical 12 and 21, 13 and 31,
	and 11 elements of the 
	\hspace*{0.75in} quark and lepton Dirac mass matrices.}
\end{figure}
\newpage
% FIG. 3
\begin{figure}
\phantom{x}
%\special{psfile=ab6_fig3.ps hoffset=-90 voffset=-200}
%\special{psfile=ab6_fig3.ps}
\centerline{
\epsfxsize=1\hsize
\epsfbox{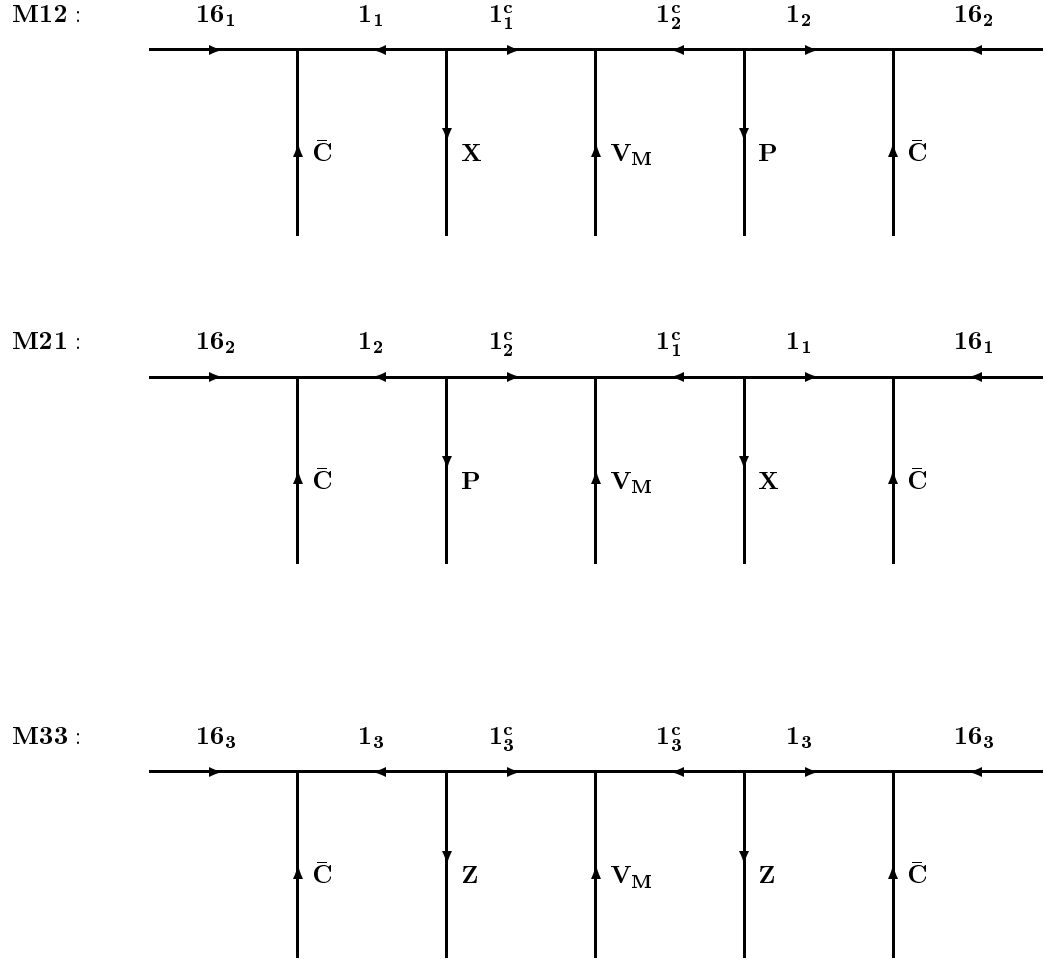}
}
\vspace{2.5in}
\caption{Diagrams that generate the 33, 12, and 21 elements of the
         Majorana mass matrix $M_R$ \hspace*{0.75in} of the superheavy 
	 right-handed neutrinos.}
\end{figure}
\newpage
% FIG. 4
\begin{figure}
\phantom{x}
%\special{psfile=ab8_fig4.ps hoffset=-90 voffset=-200}
%\special{psfile=ab8_fig4.ps}
\centerline{
\epsfxsize=1\hsize
\epsfbox{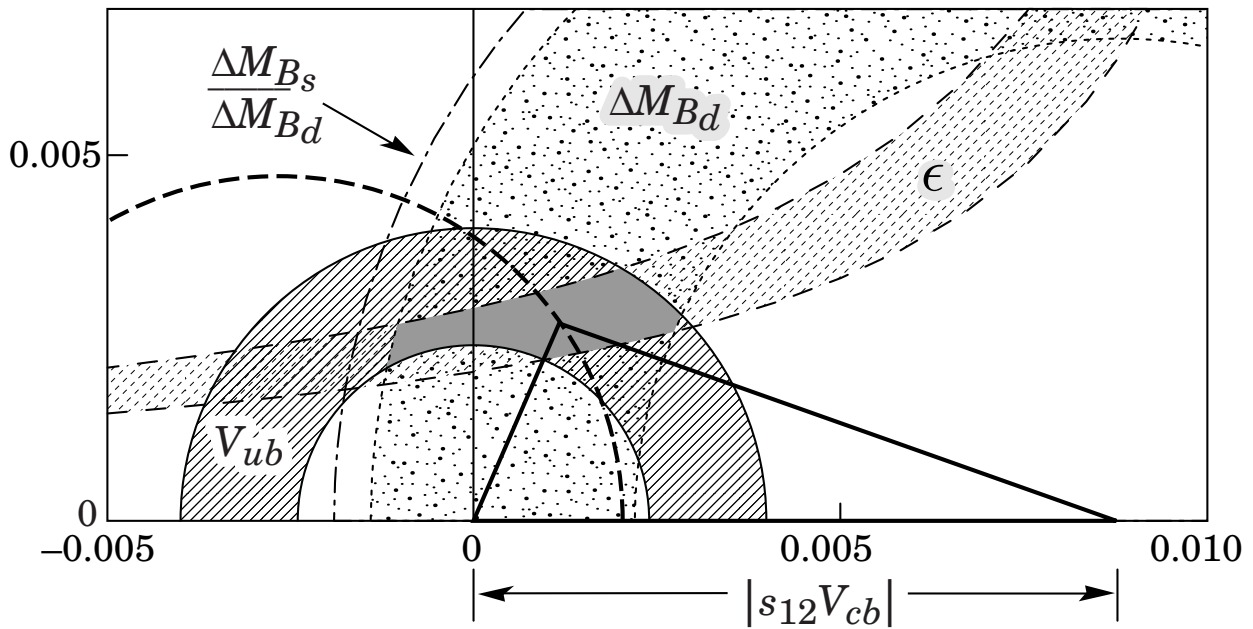}
}
\vspace{0.5in}
\caption{The unitarity triangle for $V_{ud}V^*_{ub} + V_{cd}V^*_{cb}
        + V_{td}V^*_{tb} = 0$ is displayed along with the experimental 
        constraints
        on $V_{ud}V^*_{ub}$, which is the upper vertex in the triangle.
        The constraints following from
        $|V_{ub}|$, B-mixing and $\epsilon$ extractions from experimental
        data are shown in the lightly shaded regions. The experimentally
        allowed region is indicated by the darkly shaded overlap.
        The model predicts that $V_{ud}V^*_{ub}$ will lie on the dashed
        circle. The particular point on this circle used to 
        draw the triangle shown is obtained from the CP-violating input  
        phase, $\delta_{CP} = 64^o$.}
\end{figure}
\end{document}